\documentclass[review]{elsarticle}
\usepackage{lineno,hyperref}
\usepackage{color}
\usepackage{soul}
\journal{}
\begin{document}
\begin{frontmatter}
\title{Elastic and Fracture Properties of Single Walled Pentagraphene Nanotubes}
\author[primeiro]{J. M. De Sousa\corref{author}}
\cortext[author]{Corresponding author}
\ead{josemoreiradesousa@ifpi.edu.br}
\author[mysecondaryaddress]{A. L. Aguiar}
\author[mysecondaryaddress]{E. C. Gir\~ao}
\author[mymainaddress]{Alexandre. F. Fonseca}
\author[Unicamp]{V. R. Coluci}
\author[mymainaddress]{D. S. Galv\~ao\corref{mycorrespondingauthor}}
\cortext[mycorrespondingauthor]{Corresponding author}
\ead{galvao@ifi.unicamp.br}
\address[primeiro]{Instituto Federal do Piau\'i - IFPI, S\~ao Raimundo Nonato, Piau\'i, 64770-000, Brazil}
\address[mymainaddress]{Applied Physics Department and Center of Computational Engineering and Science, University of Campinas - UNICAMP, Campinas-SP 13083
-959, Brazil.}
\address[mysecondaryaddress]{Departamento de F\'isica, Universidade Federal do Piau\'i, Teresina, Piau\'i, 64049-550, Brazil}
\address[Unicamp]{School of Technology, University of Campinas - UNICAMP, Limeira, 13484-332 SP, Brazil}

\begin{abstract}
Membranes of carbon allotropes comprised solely of densely packed pentagonal rings, known as pentagraphene, exhibit negative Poisson's ratio (auxetic behavior) and a bandgap of 3.2 eV. In this work, we investigated the structural stability, mechanical and fracture properties of nanotubes formed by rolling up pentagraphene membranes, the so-called pentagraphene nanotubes (PGNTs). Single-walled PGNT of three distinct configurations: zigzag, $\alpha$-armchair, and $\beta$-armchair were studied combining first-principles calculations and reactive molecular dynamics simulations. Our results showed Young's modulus values of 680--800 GPa, critical strain of 18--21\%, and ultimate tensile stress of 85--110 GPa. We also observed auxetic behavior. During  stretching at room temperature, we observed a transition between $\beta$-armchair to $\alpha$-armchair PGNT close to the critical strain. With relation to fracture patterns, we observed that mechanical failure starts at bonds mostly aligned to the stretching direction and after tube radial collapse.
\end{abstract}

\begin{keyword}
pentagraphene nanotubes, mechanical properties, DFT, reactive molecular dynamics, nanotechnology, fracture. 
\end{keyword}

\end{frontmatter}

\section{Introduction}

One-dimensional (1D) systems with tubular forms have attracted much attention in the scientific community in the last decades. The most representative examples are the carbon nanotubes (CNT) \cite{iijima1991helical}.  Represented as graphene sheets rolled up to form seamless cylinders, CNT can be of different chiralities; armchair, zigzag and chiral geometries \cite{saito1998physical}. Their mechanical properties, especially the high elasticity modulus, have been widely exploited in nanotechnology applications \cite{dresselhaus1995physics,ajayan2001applications}.

The search for new 1D systems with properties similar to CNT has been intense in recent years. Examples of interesting structures include graphyne nanotubes  (GNT) \cite{coluci2004theoretical,de2016torsional}, boron and nitrogen nanotubes (BNNT) \cite{chen2004boron, suryavanshi2004elastic}, carbon and nitrogen nanotubes \cite{sung1999well,guo2004synthesis}, titanate nanotubes \cite{sun2003synthesis, lan2005titanate}, and pentagraphene nanotubes (PGNT) \cite{chen2017mechanical,de2018mechanical}. Considering a single chirality of PGNT ($\alpha$-armchair), Chen \textit{et. al} predicted that these PGNT have plastic behavior under tensile strains with irreversible pentagon-to-polygon structural transformations \cite{chen2017mechanical}. It has been also demonstrated \cite{de2018mechanical} that PGNT fracture patterns depend on their chirality, and that $\alpha$-armchair-PGNT break at smaller tensile strains than zigzag ones \cite{de2018mechanical}. Despite the increasing interest in PGNT, a detailed analysis of their mechanical properties is still missing.

In this work, we investigated the mechanical properties of PGNT of different chiralities; zigzag, armchair-$\alpha$ and armchair-$\beta$. The mechanical properties under elastic deformations and at fracture conditions were investigated combining density functional theory calculations (DFT) and fully atomistic reactive molecular dynamics (MD) simulations. Our results show that the Young's modulus values predicted for the PGNT are smaller than that of conventional CNT. In particular, the PGNT-$\beta$-armchair nanostructures undergo interesting structural transitions when stretched, due to a transformation that occurs in their atomic configuration that lead them to a more stable configuration, the armchair-$\alpha$ PGNT. 

\section{Generation of Pentagraphene Nanotubes}
\begin{figure}[ht!]
\centering
\includegraphics[width=1.0\columnwidth]{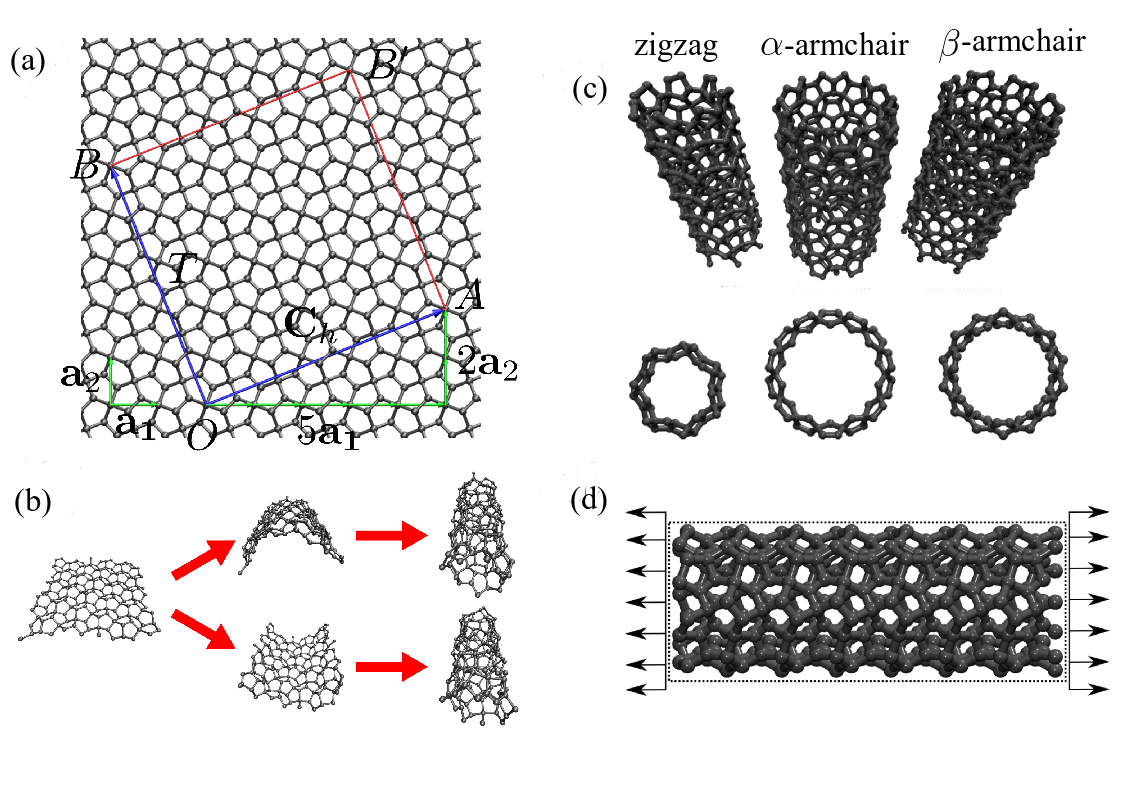}
\caption{(a) Unit cell of a pentagraphene membrane. (b) Rolling process of a pentagraphene membrane for the case of (5,2) PGNT to form the $\alpha-$(5,2) (top) and $\beta-$(5,2) (bottom).  (c) Atomic models for PGNT with different chiralities: zigzag $(7,0)$, armchair-$\alpha$ $(7,7)$ and armchair-$\beta$ $(7,7)$, respectively and a perspective view of the PGNT. (d) Representation of an axial load (along the arrows) applied into a PGNT.}
\label{fig1a}
\end{figure}

The unit cell of a pentagraphene membrane can be chosen to be a regular square with side length $a$ ($a\approx$3.64\AA) generated by the lattice vectors $\mathbf{a}_1=a(1,0)$ and $\mathbf{a}_2=a(0,1)$ (Fig.~\ref{fig1a} (a)). 
In order to build a PGNT, we define a chiral vector $\mathbf{C}_h$ in a similar way as for conventional single-walled carbon nanotube (SWCNT) as
\begin{eqnarray}
\mathbf{C}_h=(n,m)=n\mathbf{a}_1+m\mathbf{a}_2.
\end{eqnarray}

The angle between $\mathbf{C}_h$ and $\mathbf{a}_1$, the chiral angle $\theta_c$, is defined as
\begin{eqnarray}
\cos\theta_c=\frac{n}{\sqrt{n^2+m^2}}.
\end{eqnarray}
Due to the $S_4$ symmetry at the $sp^3$ carbon atom serving as origin for the $\mathbf{a}_1$ and $\mathbf{a}_2$ vectors, we can restrict the valid $\mathbf{C}_h$ vectors to those with $n,m\ge0$. In addition, symmetry relative to $\theta_c=45^\circ$ allows us to limit $m$ to the $[0,n]$ interval. The translational vector, defined as the smallest vector orthogonal to $\mathbf{C}_h$, is written as
\begin{eqnarray}
\mathbf{T}=(t_1,t_2)=t_1\mathbf{a}_1+t_2\mathbf{a}_2,
\end{eqnarray}
with $t_1$ and $t_2$ being integers. Using the fact that $\mathbf{C}_h\cdot\mathbf{T}=0$ and choosing $t_2$ to be positive, we obtain
\begin{eqnarray}
t_1=-m/d\quad\textrm{and}\quad t_2=n/d,
\end{eqnarray}
where $d$ is the maximum common divisor of $n$ and $m$. In Fig.~\ref{fig1a} (a) we illustrate the $\mathbf{C}_h=(5,2)$ vector and its corresponding $\mathbf{T}=(-2,5)$ counterpart. The number of atoms in the tube unit cell is given by six times the number $\mathcal{N}$ of $a\times a$ squares within the rectangle defined by $\mathbf{C}_h$ and $\mathbf{T}$. The $\mathcal{N}$ value is the ratio of the norms of the vectors $\mathbf{C}_h\times\mathbf{T}$ and $\mathbf{a}_1\times\mathbf{a}_2$, resulting in
\begin{eqnarray}
\mathcal{N}=\frac{|\mathbf{C}_h\times\mathbf{T}|}{|\mathbf{a}_1\times\mathbf{a}_2|}=\frac{(m^2+n^2)}{d}.
\end{eqnarray}
The length $L$ and radius $R$ of the PGNT are given in terms of $\mathbf{T}$ and $\mathbf{C}_h$, respectively, by
\begin{eqnarray}
L=|\mathbf{T}|=\frac{\sqrt{n^2+m^2}}{d}a\quad\textrm{and}\quad R=\frac{|\mathbf{C}_h|}{2\pi}=\frac{\sqrt{n^2+m^2}}{2\pi}a.
\end{eqnarray}

Similarly to the single-walled CNT case, the PGNT unit cell is obtained by rolling up the rectangle sector of the membrane determined by $\mathbf{C}_h$ and $\mathbf{T}$, joining the points $O$ to $A$ and $B$ to $B'$. However, there are two ways to roll up the membrane by folding it either up or down, as illustrated in Fig.~\ref{fig1a} (b). Differently from CNT, choosing any of these folding directions results in different nanotube configurations. This is due to the thickness of the membrane and the arrangement of the carbon bonds.  We can see from the Fig. that the carbon-carbon bonds between two tri-coordinated atoms can be grouped into two sets. All the bonds in a given set are simultaneously parallel to each other and orthogonal to the corresponding bonds from the other set. Thus, choosing the folding direction means to choose which set of bonds will be in the outer wall of the tube (the other set going to the inner part of the tube wall). We identify the two different tube configuration corresponding to a given $(n,m)$ as $\alpha-(n,m)$ and $\beta-(n,m)$. According to our used notation, $\alpha$ PGNT appear when we fold the membrane in the direction of the $\mathbf{T}\times\mathbf{C}_h$ (top case in Fig.~\ref{fig1a} (b)), whereas $\beta$ PGNTs appear when we fold the membrane in the direction of the $\mathbf{C}_h\times\mathbf{T}$ (bottom case in Fig.~\ref{fig1a} (b)). Atomistic representations of PGNT generated by this process are shown in Fig. \ref{fig1a} (c).

\section{Methods}
To determine the PGNT structural stability and their elastic properties, we carried out a systematic study using a DFT framework\cite{hohenberg64,Kohn65} as implemented in the SIESTA code\cite{ordejon96,portal97}. The Kohn-Sham orbitals were expanded in a double-$\zeta$ basis set composed of numerical pseudoatomic orbitals of finite range enhanced with polarization orbitals. A common atomic confinement energy shift of 0.02 Ry was used to define the basis function cutoff radii, while the fineness of the real space grid was determined by a mesh cutoff of 400 Ry\cite{anglada02}. For the exchange-correlation potential, we used the generalized gradient approximation\cite{perdew96} and the pseudopotentials were modeled within the norm-conserving Troullier-Martins \cite{troullier91} scheme in the Kleinman-Bylander \cite{kleinman82} factorized form. Brillouin-zone integrations were performed using
a Monkhorst–Pack\cite{monkhorst76} grid of 1 $\times$ 1 $\times$ 8 $k$-points. Periodic boundary conditions were imposed, with perpendicular lattice vectors a$_x$ and a$_y$ large enough ($\sim$ 40\AA{}) to simulate vacuum and to avoid spurious interactions among periodic images. 

The PGNT structural stability was estimated by calculating the formation energy per atom $\varepsilon_f$, as given by $E_{PGNT}/N_{PGNT}-$ $\varepsilon_{PG}$, where $E_{PGNT}$ is the total energy, $N_{PGNT}$ is the number of atoms of simulated PGNT, and $\varepsilon_{PG}$ is the total energy per atom of the pentagraphene membrane. 

The strain energy was obtained after PGNT stretching. The strain was calculated as $\varepsilon_{z}=L/L_0$, where $L_0$ and $L$ are the relaxed and strained tube length, respectively. For each strain value, the PGNT nanostructure was fully relaxed until the maximum force component on each atom was less than 0.01 eV/\AA. For each structural relaxation, the SCF convergence thresholds for the electronic total energy were set at 10$^{-4}$ eV. 

To determine the elastic properties, the PGNT were treated as a rolled membrane with thickness $h$ equal to 3.64 \AA{}  and area equal to $\pi d L_0$ where $d$ is the tube diameter. Therefore, the axial stress component $\sigma_{z}$ is related to strain component $\varepsilon_{z}$ as $\sigma_{z}=(1/\Omega)(\partial U/\partial\varepsilon_{z})$
where $\Omega=L_0\pi d_t h$ is the volume of the former tube membrane. Young's modulus, $Y$, is estimated from the slope ($d\sigma/d\varepsilon$) of strain-stress curves in the linear regime.

In order to estimate the stretching behavior of PGNT at room temperature, we carried out MD simulations with the ReaxFF force field \cite{mueller2010development, van2001reaxff}. ReaxFF is a force field that was developed to precisely handle large deformations/fractures and/or chemical reactions
\cite{van2001reaxff}. Its
parameterization is obtained directly from first-principles calculations and compared with experimental values. We used the LAMMPS code \cite{plimpton1995fast} to integrate the equations of motion and obtain the atomic trajectories.  The criteria of validation of the ReaxFF parameters for carbon consist of obtaining deviations between the simulated 
and experimental values for the heat of formation. For unconjugated and conjugate carbon systems these differences are not larger than about $2.8$ and $2.9$ kcal/mol, respectively \cite{van2001reaxff}. The main characteristic of ReaxFF is the description in the formation/breaking of chemical bonds as a function of the bond order values, thus being ideal for the study of the mechanism of fracture in nanostructured systems.
In ReaxFF, the energy of the system is divided into partial energy contributions as shown in Eq. \ref{Eq1} \cite{van2001reaxff}:
\begin{eqnarray}
E_{system} &=& E_{bond} + E_{over} + E_{under} + E_{val} + E_{pen} +\nonumber \\
&& E_{tors} + E_{conj} + E_{vdW} + E_{Coul}
\label{Eq1}
\end{eqnarray} 
where the partial energy contributions of the system are: covalent bond  $(E_{bond})$, excess of bonds $(E_{over})$, absence of bond $(E_{under})$, valence angle $(E_{val})$, penalizations $(E_{pen})$, torsion $(E_{tors})$, conjugated systems $(E_{conj})$, and no bonded interactions like van der Waals, $(E_{vdW})$ and Coulomb interactions $(E_{Coul})$ \cite{van2001reaxff}.

For MD simulations at room temperature, we considered PGNT of several diameters and chiralities (zigzag, $\alpha$-armchair and $\beta$-armchair) with periodic boundary conditions along with the axis $z$. For zigzag $(n,0)$ PGNT, we considered tube lengths of 2.54 nm (7 unit cells) and diameters ranging from 0.579 nm ($n=5$) to 1.390 nm ($n=12$), which corresponds to 210 and 504 carbon atoms, respectively. In the case of $\alpha$-armchair and $\beta$-armchair $(n,n)$ PGNT, we considered tube length of of 2.57 nm (5 units cells) and diameters ranging from 0.819 nm ($n=5$) to 1.966 nm ($n=12$), which corresponds to 300 and 720 carbon atoms, respectively, in the simulation box.

To eliminate initial stress before the stretching process, we thermalized the nanotubes within the isothermal-isobaric ensemble \cite{evans1983isothermal}, setting the pressure to zero. In all MD simulations, the temperature was kept constant at 300 K and was controlled by a Nose-Hoover thermostat \cite{hoover1985canonical}.

The stretching was produced by increasing the simulation box size along the periodic direction (the $z$-axis). The dynamics of our system was updated for each increment of 0.05 fs, when the deformation rate was set constant and equals to $10^{-6}$ fs$^{-1}$. The elastic properties were characterized by the Young's modulus, obtained as $Y = 
d\sigma_{ii}/d\varepsilon_{i}$,
where $\sigma_{ii}$ is the component of the virial tensor stress and $\epsilon_{i}$ is the deformation along the axial direction $i$. The stress tensor is defined as:
\begin{eqnarray}
\sigma_{ij} = \frac{\sum_{k}^{N}m_{k}v_{k_{i}}v_{k_{j}}}{A} + \frac{\sum_{k}^{N}m_{k}r_{k_{i}}.f_{k_{j}}}{A},
\label{Eq3}
\end{eqnarray}
where $A$ is the area of the PGNT, $N$ is the number of carbon atoms, $v$ is the velocity, $r$ is the carbon atoms positions and $f$ is the force per atom. The spatial stress distribution per atom is calculated using the von Mises stress tensor\cite{botari2014mechanical,de2016mechanical,bizao2018scale,woellner2018structural,de2019elasticGNTs}, defined as
\begin{eqnarray}
\sigma_{VM}^{i} = \sqrt{\frac{(\sigma_{xx}^{i}-\sigma_{yy}^{i})^{2} + 
(\sigma_{yy}^{i}-\sigma_{zz}^{i})^{2} + (\sigma_{xx}^{i}-\sigma_{zz}^{i})^{2} +
6 \left[(\sigma_{xy}^{i})^{2} + (\sigma_{yz}^{i})^{2}  + (\sigma_{zy}^{i})^{2}\right]}{2}} .
\label{Eq4}
\end{eqnarray} 

\section{Results}

\subsection{Structural stability, bond distribution, and elastic properties}

\begin{figure}[ht!]
\centering
\includegraphics[width=0.50\columnwidth]{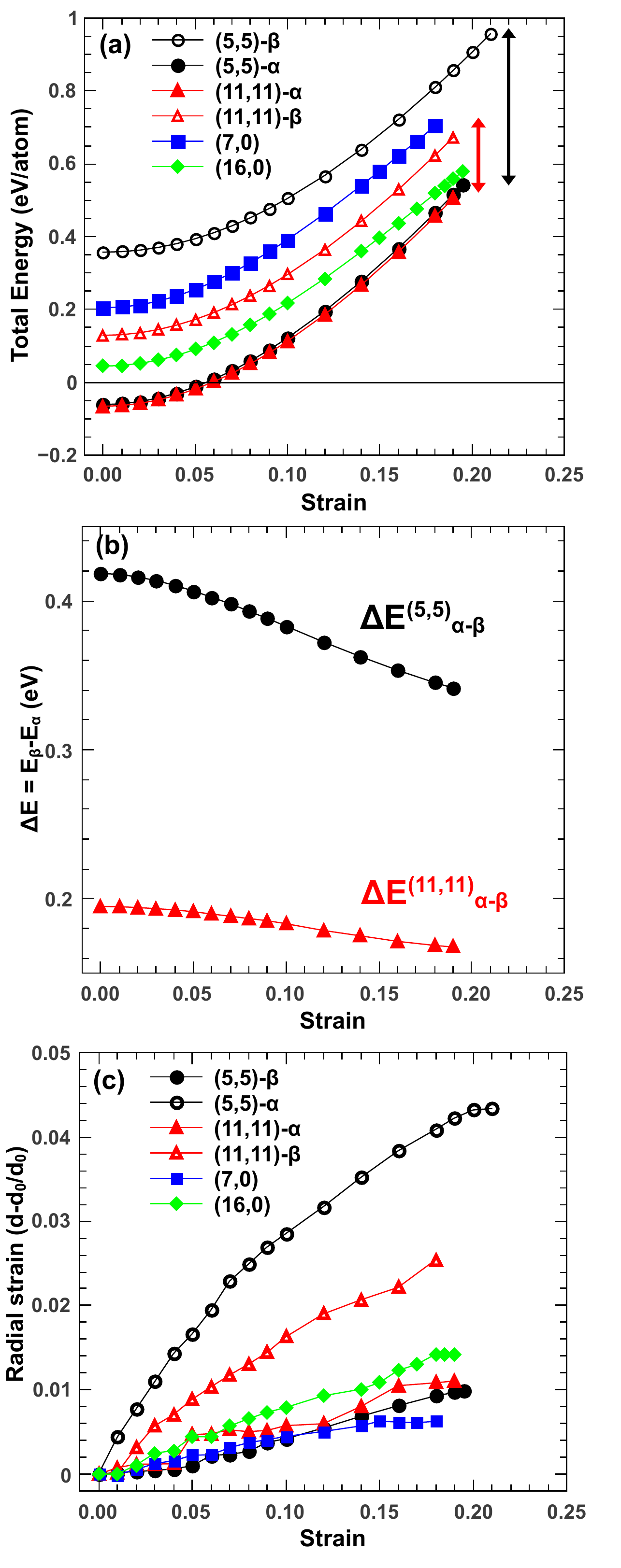}
\caption{(a) PGNT total energy as a function of the applied tensile strain. (b) Energy difference $\Delta E$ between $\beta$-armchair and $\alpha$-armchair PGNT as a function of strain. (c) Radial strain of PGNT as a function of axial strain. The radial strain is defined as $\varepsilon_r = (d-d_0)/d_0$, where $d_0$ is the average diameter for unstretched PGNT.}
\label{energy}
\end{figure}

In Fig. \ref{energy} we present DFT results for tube energy configurations as a function of the applied strain values. As we can see from Fig. \ref{energy}a, $\alpha$-armchair PGNT are the only nanotubes with negative total energy values in the unstretched state. When stretched, the strain energy of the $\alpha$-armchair PGNT ($\alpha$-(11,11) and $\alpha$-(5,5)) show no significant dependence on diameter values. Furthermore, $\alpha$-armchair PGNT are energetically more favorable than $\beta$-armchair ones, for all the stretched cases analyzed here (Fig. \ref{energy}b). As expected, the energy difference $\Delta E$ between $\beta$-armchair and $\alpha$-armchair PGNT is smaller for the larger diameter considered here. This difference is significantly reduced with the applied strain values for the (5,5)-armchair PGNT. 

We have observed that, when stretched, all the studied PGNT displayed a small (up to 5\%) increase on the average diameter $d$ values, as compared to the initial average diameter $d_0$ (Fig. \ref{energy}c). This increase indicates that all the studied PGNT have negative Poisson's ratio $\nu=-\varepsilon_{r}/\varepsilon_{z}$, similar to what happens for the pentagraphene membrane. $\beta$-armchair PGNT exhibit higher absolute $\nu$ values than $\alpha$-armchair and zigzag PGNT. Poisson ratio values estimated close to equilibrium for (5,5) $\beta$-PGNT and (11,11) $\beta$-PGNT were $\nu=-0.331$ and $\nu=-0.193$, respectively, whereas for (5,5) $\alpha$-PGNT and (11,11) $\beta$-PGNT were $\nu=-0.018$ and $\nu=-0.071$,
respectively. Smaller values were found for zigzag PGNT, namely $\nu=-0.048$ for (7,0) and $\nu=-0.091$ for (16,0). Whereas the thickness of the unstretched (5,5)$\beta$-PGNT is higher than that for the (5,5)$\alpha$-PGNT, both  stretched $\alpha$-PGNT and $\beta$-PGNT  converged to a smaller but very similar thickness. Similar results were found for (11,11)$\alpha$-PGNT and (11,11)$\beta$-PGNT. For (7,0) and (16,0) zigzag PGNT, their thickness is also significantly reduced as the nanotubes are stretched.

\begin{figure}[htb!]
 \centering
 \includegraphics[scale=0.30]{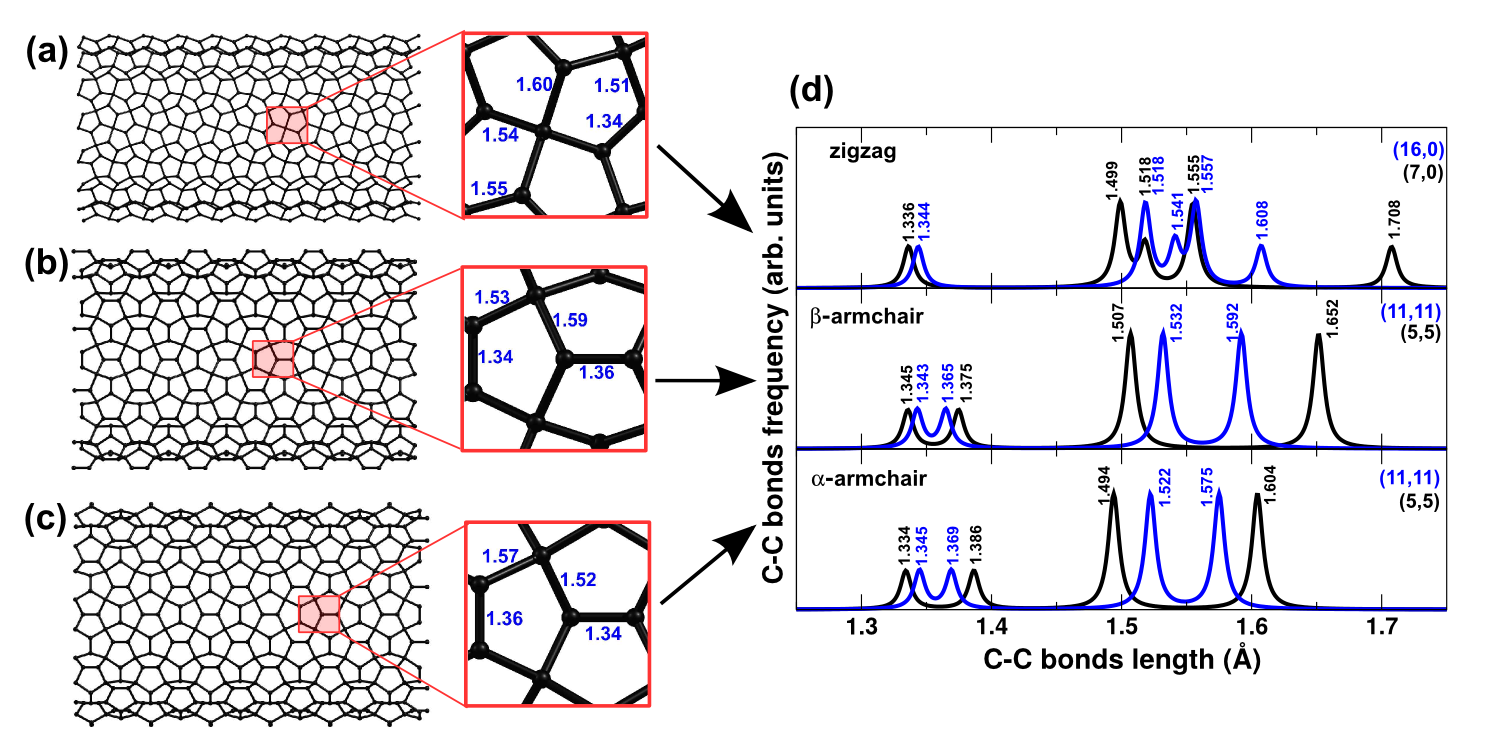}
 \caption{Analysis of the $C-C$ bond-length values of the PGNT optimized with DFT.
  The values obtained for (16,0) PGNT (a), $\beta$-(11,11) PGNT (b), $\alpha$-(11,11) PGNT (c) are shown in the zoomed regions.
  (d) Bond-length values distribution for (16,0) and (7,0) PGNT (top),
  $\beta$-(11,11) PGNT and $\beta$-(5,5) PGNT (middle), and $\alpha$-(11,11) PGNT and $\alpha$-(5,5) PGNT (bottom).}
\label{fig_bonds}
\end{figure}

DFT bond-length values for the optimized nanostructures of the (16,0) zigzag, $\beta$-(11,11) armchair, and $\alpha$-(11,11) armchair PGNT are shown in Fig. \ref{fig_bonds}a-c. Significant differences can be identified for the bond lengths of zigzag and armchair PGNT. Within the representative nanotube portion (panels in Fig. \ref{fig_bonds}), only a single bond with a bond length in the 1.30--1.40 \AA  range (single C--C bond) was observed for zigzag PGNT, whereas two single bonds were observed for both $\alpha$- and $\beta$-armchair PGNT. We identified four bonds within the 1.50--1.70 \AA  range for zigzag PGNT and two of them for $\alpha$- and $\beta$-armchair PGNT. The length values of those bonds are more dependent on the nanotube diameter than the single bond-lenghts (Fig. \ref{fig_bonds}d). Our values for the bond lengths are in good agreement with the values obtained for pentagraphene membranes \cite{zhang2015penta}. The C-C bonds of $\beta$-armchair PGNT are clearly stretched and diameter dependent when compared to $\alpha$-armchair PGNT, which agrees with the energy stability values found for $\alpha$-armchair PGNT. 

\begin{figure}[ht!]
\centering
\includegraphics[width=1.05\columnwidth]{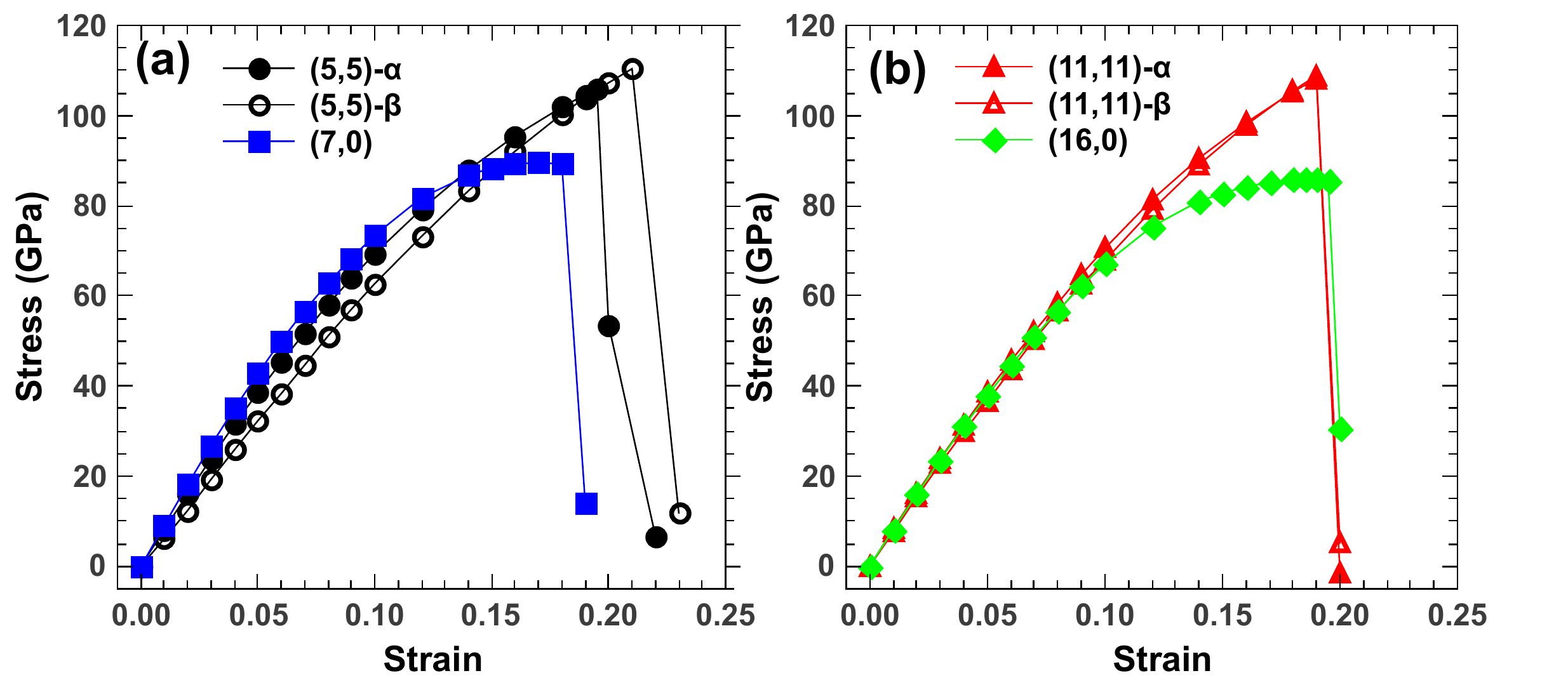}
\caption{Stress-strain curves of PGNTs predicted by DFT calculations.}
\label{strain}
\end{figure}

DFT results for the stress-strain curves for PGNT indicate a slightly difference between the small diameter (5,5)$\alpha$-armchair and  (5,5)$\beta$-armchair PGNT (Fig. \ref{strain}a) and no significant differences for (11,11)$\alpha$ and (11,11)$\beta$ PGNT (Fig. \ref{strain}a,b).  Critical strains of 18--21\% were obtained for all the studied systems. These values are similar to the ones predicted for the pentagraphene membrane \cite{de2017mechanical}. The ultimate tensile stress were 85--90 GPa for zigzag PGNT and 105--110 GPa for armchair PGNT. After the fracture of the $(7,0)$ and $(16,0)$ PGNT, we have observed 
the formation of large regions of the stable carbon phase biphenylene  (not shown), which has been proposed as the resulting configuration of  pentagraphene membrane after fracture \cite{rahaman2017metamorphosis}.

\begin{figure}[htb!]
 \centering
 \includegraphics[scale=0.8]{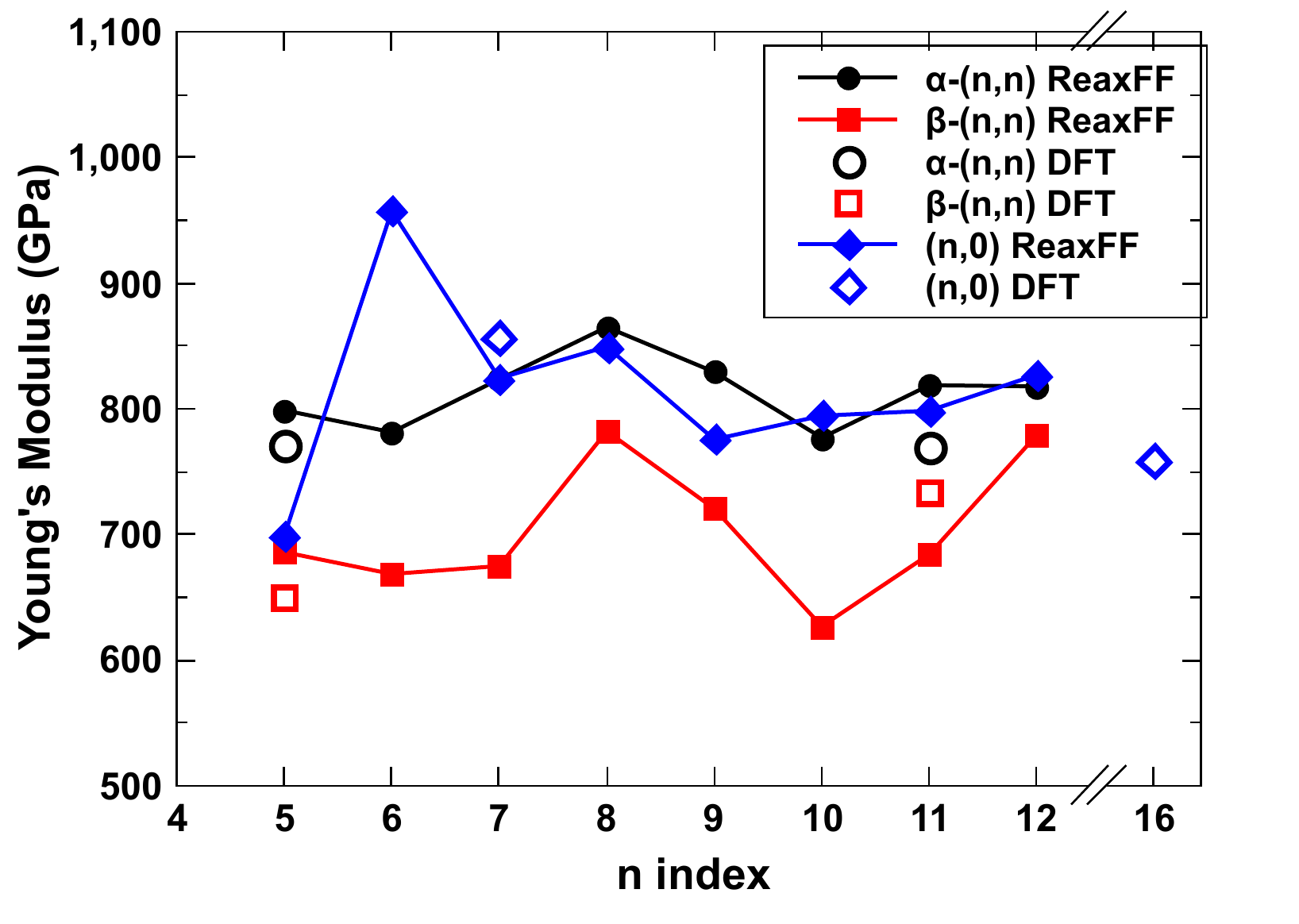}
 \caption{PGNT Young's modulus values as function of the $n$ tube index.}
\label{ym}
\end{figure}

The Young's modulus values for all the studied PGNT are in the range of 600--1000 GPa (Fig. \ref{ym}). Predictions from MD simulations are in good agreement with DFT results. We found no significant dependence of $Y$ on the nanotube diameter value (which is proportional to the $n$ index). However, we can clearly see that the Young's modulus values of PGNT-$\beta$-armchair are smaller than that of PGNT-zigzag and PGNT-$\alpha$-armchair. Thus, the MD results suggest that this difference, as discussed above, is due to its atomic configuration, where the directions of the bonds are aligned with the direction of the uniaxial strain, influencing the accumulated stress distribution in the PGNT, and thus altering the fracture patterns in the PGNT. From DFT calculations, we have also observed that the diameter and thickness of PGNT-$\beta$-armchair are more susceptible than the other PGNT studied, and this could explain why PGNT-$\beta$-armchair are more easy to stretch than the others.

\subsection{Fracture patterns}

In order to get insights into the PGNT fracture patterns at room temperature, when they are close to yield strain,
we carried out MD simulations of the nanotubes with large supercells. The ReaxFF potential was first tested by analyzing the lengths of the C--C bonds on the different PGNT and comparing them to the predictions from DFT calculations depicted in 
Fig. \ref{fig_bonds}. These bond lengths assumed values between 1.53 \AA and 1.57 \AA close to the ones predicted by DFT calculations (between 1.535 \AA and 1.555 \AA) \cite{zhang2015penta}. 

\begin{figure}[htb!]
 \centering
 \includegraphics[scale=0.30]{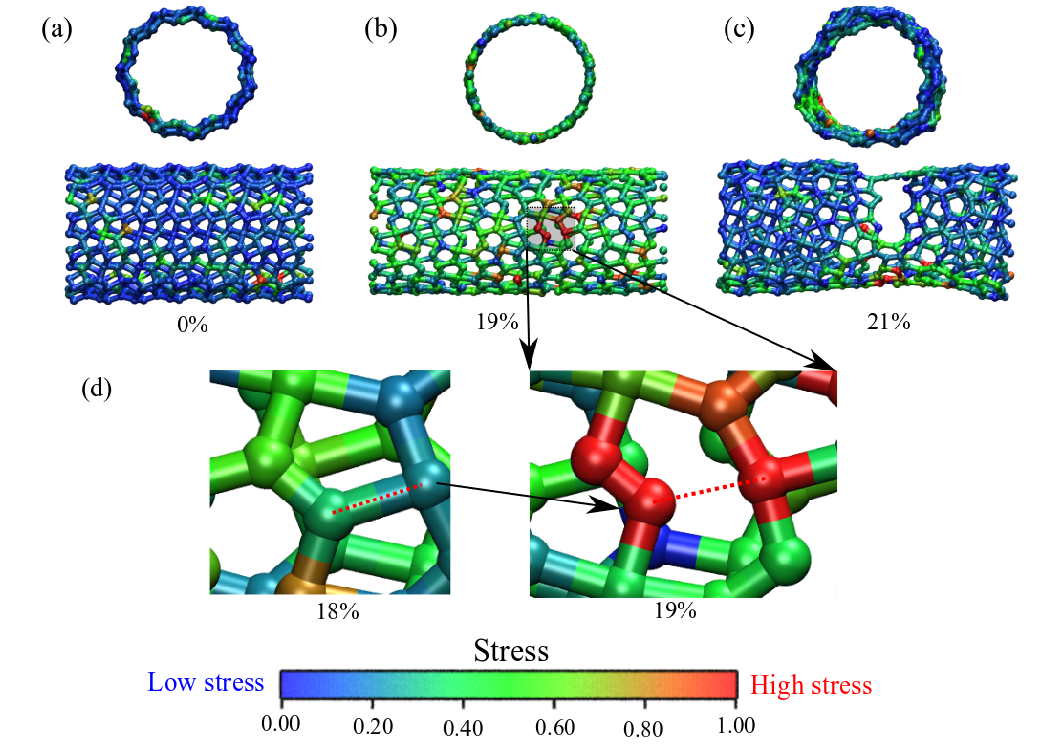}
 \caption{Atomic models for PGNT-zigzag $(11,0)$ at different strain conditions: 
 unstretched, (b) 19 \% and (c) 21 \%. A process of breaking one bond (red dotted line) at 19 \% is shown in (d). 
 The horizontal bar represents the level of von mises stress, where the color blue (red) represents 
low (high) stress.}
\label{fig_PGNTzigzag}
\end{figure}

\begin{figure}[htb!]
 \centering
 \includegraphics[scale=0.30]{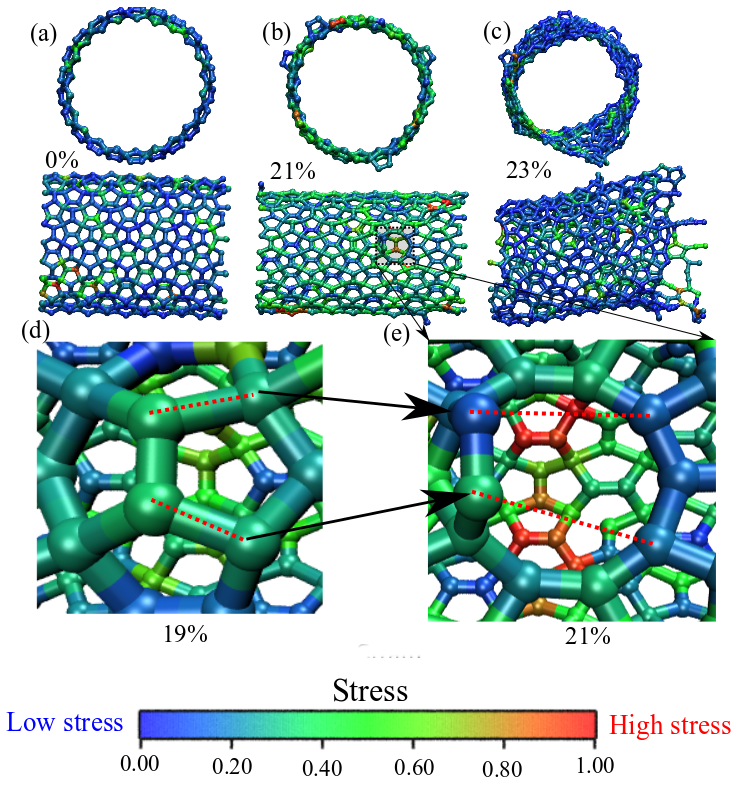}
 \caption{Atomic models for $\alpha$-armachair PGNT (11,11) at different strain conditions: 
 (a) 
 unstretched, (b) 21 \% and (c) 23 \%. A process of breaking one bond (red dotted line) at 21 \% is shown in (d).}
\label{fig_PGNT_armc_alfa}
\end{figure}

\begin{figure}[htb!] 
 \centering
 \includegraphics[scale=0.30]{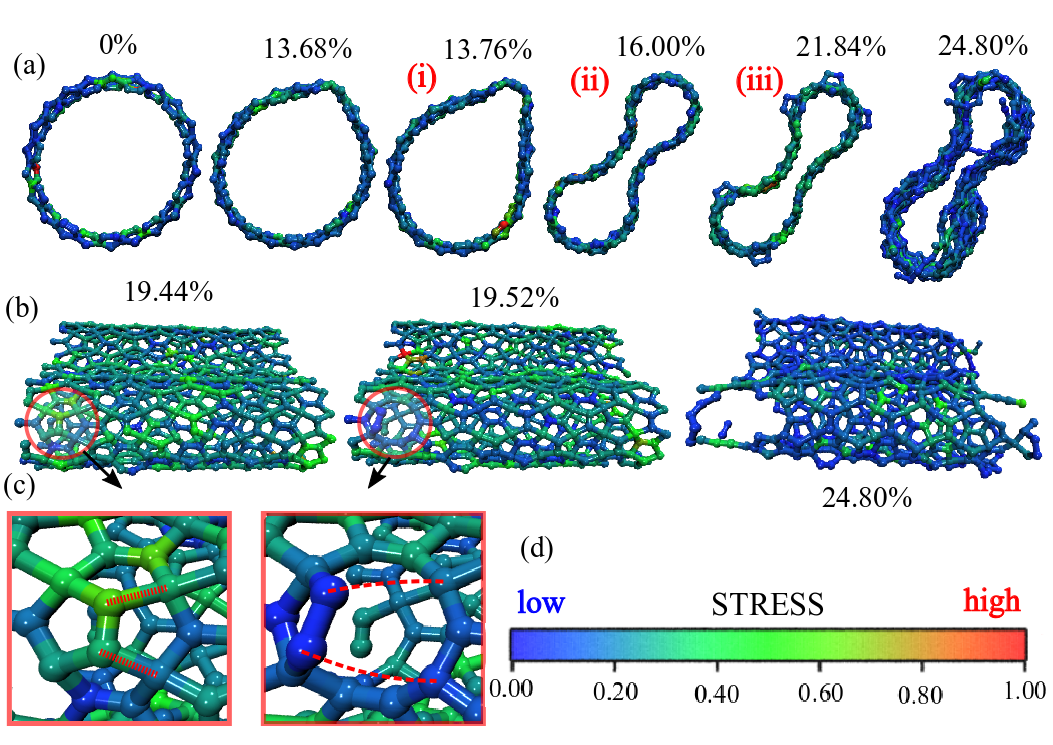}
 \caption{(a) Cross sections of armchair PGNT (11,11) at different strain conditions . The structural transition from the $\beta$-armchair to $\alpha$-armchair is indicated in (i)($\beta$), (ii)($\beta$) and (iii)($\alpha$). (b)-(c) Fracture pattern of the armchair (11,11) during the transition.}
\label{fig_PGNT_armc_beta}
\end{figure}

The stretching evolution showed similar fracture patterns for zigzag and armchair PGNT (Figs. \ref{fig_PGNTzigzag}, \ref{fig_PGNT_armc_alfa} \ref{fig_PGNT_armc_beta}). MD snapshots of the evolution of the (11,0) PGNT for different strains show the fracture occurs between 19\% and 21\% strain (Fig. \ref{fig_PGNTzigzag}). The fracture starts at the bonds that are approximately aligned with the direction of the applied strain (dashed lines in Fig. \ref{fig_PGNTzigzag} d). Similarly to zigzag PGNT, the bond-breaking also occurs for bonds aligned to the strain direction (Fig. \ref{fig_PGNT_armc_alfa}). This is illustrated in Fig. \ref{fig_PGNT_armc_alfa}d for the (11,11)-$\alpha$-armchair PGNT when two bonds (dashed lines) from a single pentagon break at once. This fracture pattern was observed to occur at several places of the nanotube. 
In Figs. \ref{fig_PGNT_armc_beta} (b) and (c) we can observe the mechanical fracture pattern of PGNT-$\beta$-armchair, which is similar to that of PGNT-$\alpha$-armchair. 

\begin{figure}[htb!]
 \centering
 \includegraphics[scale=1.35]{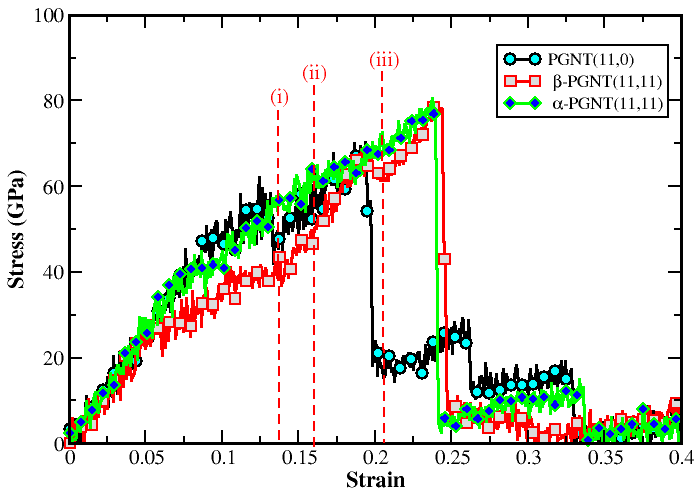}
 \caption{Stress versus strain curves for (11,0) zigzag PGNT (black), (11,11) $\alpha$-armchair PGNT (green) and (11,111) $\beta$-armchair PGNT (red) predicted by MD simulations with the ReaxFF  potential at 300 K. The points $(i)$, $(ii)$ and $(iii)$ represent the nanostructures depicted in Fig. \ref{fig_PGNT_armc_beta}(a).}
\label{stress_strain}
\end{figure}

The stretching of the (11,11) $\beta$-armchair at room temperature revealed a structural transition from the $\beta$ to $\alpha$-armchair (Fig. \ref{fig_PGNT_armc_beta}). As predicted by DFT calculations (Fig. \ref{energy} b), (11,11) $\alpha$-armchair has a formation energy 0.16 eV/atom smaller than (11,11) $\beta$-armchair. Thermal activation and, more importantly,  stretching induced the transition at a strain of 16\% and a fracture at a critical strain of about 25\%. This critical strain is slightly larger than the one for the $\alpha$-armchair (23\%) and larger than the one for the zigzag (11,0) PGNT (21\%).  The three configurations (i), (ii) and (iii), shown in Fig. \ref{stress_strain} for the $\beta$--armchair PGNT, correspond to the initial configuration, the configuration at the beginning of transition from $\beta$-armchair to $\alpha$--armchair, and 
the complete fracture of $\beta$--armchair, already converted to a $\alpha$--armchair, respectively. We attributed this difference to the nanotube collapse within the plastic regime (Fig. \ref{fig_PGNT_armc_beta} a -(i),(ii), (iii) and Fig. \ref{stress_strain}) which turns out to be an intermediate step before the complete fracture and to the atomic density. The (11,11) $\alpha$--armchair presents approximately 30\% more atoms in its unit cell than the (11.0) zigzag, thus there are more bonds to be broken before complete fracture, which influences the critical strain.

\section{Conclusions}

In summary,  
we have investigated the structural stability, elastic properties, and fracture patterns of single-walled pentagraphene nanotubes (PGNT) of different chiralities using fully atomistic reactive molecular dynamics simulations and first-principles methods.
Our results show that only $\alpha$-armchair PGNT present negative total energy values in the unstretched state. The PGNT present Young's  modulus values about 20\% lower 
than the ones found for conventional carbon nanotubes ($\sim$1 TPa). Critical strain values were between 18--21\%, and  ultimate tensile stress of 85--90 GPa for zigzag PGNT and 105--110 GPa for armchair PGNT. The PGNT are also predicted to exhibit auxetic behavior with negative Poisson's ratio of --0.3 for the (5,5) $\beta$-armchair PGNT. Fracture analysis revealed similar patterns for zigzag and armchair PGNT with the fracture starting at bonds mostly aligned to the stretching direction. Finally, a thermal activated, stretch induced transition was observed from $\beta$-armchair to $\alpha$-armchair PGNTs during tensile load on the (11,11) $\beta$-armchair PGNT.

\section{Acknowledgements}

This work was supported in part by the Brazilian Agencies CAPES, CNPq and FAPESP. 
J.M.S and D.S.G. thank the Center for Computational Engineering and Sciences at Unicamp for financial support through the S\~{a}o Paulo Research Foundation (FAPESP)/CEPID grant $\#$2013/08293$-$7. 
A.F.F. is a fellow of the Brazilian Agency CNPq ($\#$302750/2015$-$0) and acknowledges support from FAPESP grant $\#$2018/02992$-$4.
V.R.C. acknowledges the financial support of FAPESP (Grant $16/01736-9$). A.L.A acknowledges CENAPAD-SP for computer time and the
Brazilian agencies CNPq (grants 427175/2016-0 and 313845/2018-2) for financial
support. E.C.G. acknowledges support from CNPq (Process No. $307927/2017-2$) and Coordena\c c\~ao de Aperfei\c coamento de Pessoal de N\'ivel 
Superior (CAPES) through the Science Without Borders program (Project Number $A085/2013$). The authors J.M.S, A.L.A and E.C.G thank the Laborat\'orio de Simula\c c\~ao Computacional Caju\'ina ($LSCC$) at Universidade Federal do Piau\'i for computational support.

\newpage
\section{References}
\bibliography{PGNT.bib}
\end{document}